\documentclass{article}

\usepackage{arxiv}

\usepackage[utf8]{inputenc} 
\usepackage[T1]{fontenc}    
\usepackage{hyperref}       
\usepackage{url}            
\usepackage{booktabs}       
\usepackage{amsfonts}       
\usepackage{nicefrac}       
\usepackage{microtype}      
\usepackage{lipsum}		
\usepackage{graphicx}
\usepackage{natbib}
\usepackage{doi}
\usepackage{makecell}

\title{Comparison of Methods that Combine Multiple Randomized Trials to Estimate Heterogeneous Treatment Effects}

\author{
 Carly Lupton Brantner \\
  Department of Biostatistics\\
  Johns Hopkins Bloomberg School of Public Health\\
  \texttt{clupton1@jhu.edu} \\
   \And
 Trang Quynh Nguyen \\
 Department of Mental Health\\
 Johns Hopkins Bloomberg School of Public Health
  \And
  Tengjie Tang \\
  Department of Statistical Science\\
  Duke University
  \And
  Congwen Zhao \\
  Department of Biostatistics and Bioinformatics\\
  Duke University
  \And
  Hwanhee Hong\\
  Department of Biostatistics and Bioinformatics\\
  Duke University
  \And
  Elizabeth A. Stuart \\
  Department of Biostatistics\\
  Johns Hopkins Bloomberg School of Public Health
 \\
}

\renewcommand{\shorttitle}{\textit{Combining Trials to Estimate HTE}}

\hypersetup{
pdftitle={A template for the arxiv style},
pdfsubject={q-bio.NC, q-bio.QM},
pdfauthor={David S.~Hippocampus, Elias D.~Striatum},
pdfkeywords={First keyword, Second keyword, More},
}

\begin{document}
\maketitle

\begin{abstract}
	Individualized treatment decisions can improve health outcomes, but using data to make these decisions in a reliable, precise, and generalizable way is challenging with a single dataset. Leveraging multiple randomized controlled trials allows for the combination of datasets with unconfounded treatment assignment to better estimate heterogeneous treatment effects. This paper discusses several non-parametric approaches for estimating heterogeneous treatment effects using data from multiple trials. We extend single-study methods to a scenario with multiple trials and explore their performance through a simulation study, with data generation scenarios that have differing levels of cross-trial heterogeneity. The simulations demonstrate that methods that directly allow for heterogeneity of the treatment effect across trials perform better than methods that do not, and that the choice of single-study method matters based on the functional form of the treatment effect. Finally, we discuss which methods perform well in each setting and then apply them to four randomized controlled trials to examine effect heterogeneity of treatments for major depressive disorder.
\end{abstract}

\keywords{treatment effect heterogeneity, combining data, personalized medicine, machine learning}

\maketitle

\section{Introduction}

When tailoring treatment regimens to individual patients, one must strive to understand how different treatment options might affect the specific patient based on their characteristics or context. Rather than using a one-size-fits-all approach, clinicians and researchers are turning more towards personalized medicine with the goal of improving clinical outcomes. In this setting, the focus of estimation becomes conditional average treatment effects, i.e., how well the treatment is expected to work conditional on the person's known characteristics.

The benchmark for estimating treatment effects in an unbiased manner is most often a randomized controlled trial (RCT). In an RCT, participants are randomly assigned to treatment or control, therefore ensuring unconfounded treatment assignment and unbiased treatment effect estimates in the given sample. However, these trials often have sample sizes that are large enough to detect main effects but lack power to estimate heterogeneous treatment effects \citep{fleiss2011design} and might not be representative of a broader population. To overcome these specific issues, researchers have started combining information from multiple studies to improve treatment effect estimation. Multiple studies allow for larger sample sizes and at times a more representative sample of the target population. In the setting with multiple RCTs, meta-analysis or hierarchical models are common techniques to combine studies and estimate treatment effects. \citep{debray_get_2015, seo_comparing_2021} These approaches often do not explicitly target conditional average treatment effects though, and often only use aggregate-level data which makes it challenging to estimate treatment effects conditional on individual-level characteristics. Furthermore, meta-analysis is commonly applied within a parametric framework, which is highly interpretable but requires prespecification of effect moderators and distributional assumptions for parameters. Non-parametric approaches are worth exploring in this setting because they allow for high levels of flexibility in outcome and treatment effect functions. Relationships between covariates and treatment effect can be complex and non-linear in reality, and non-parametric machine learning methods can better handle those scenarios.

Many non-parametric approaches exist to estimate heterogeneous treatment effects \citep{kunzel2019metalearners, athey_generalized_2019, green_modeling_2012, kennedy_optimal_2020, nie_quasi-oracle_2021, dandl2022makes}; however, these approaches have generally been developed only for the single-study setting. Several of the common approaches are discussed in the section to follow (\ref{SSL}), and we subsequently extend these methods for use in multiple studies. Recent research has investigated a few non-parametric approaches for the multiple study setting, mostly geared towards combining data from one RCT with a large observational dataset \citep{yang_elastic_2020,yang_improved_2022,kallus_removing_2018,rosenman_combining_2020}. In that work, the focus is often on estimating the bias present in the observational data to determine the level at which the observational study estimates can be combined with the RCT estimates. These methods are therefore not as straightforward to use in the multiple RCT setting. With multiple RCTs, each individual trial has the benefit of unconfounded treatment assignment, but significant cross-trial heterogeneity could still exist due to both observed and unobserved factors. The focus in this case is no longer de-biasing one of the datasets, but instead determining the amount of heterogeneity present and how to account for it. 

Brantner and colleagues (\citeyear{brantner2023review}) wrote a comprehensive review of methods geared towards combining datasets to estimate treatment effect heterogeneity. That review included approaches for multiple RCTs; the most common were individual participant-level data one-stage meta-analyses \citep{debray_get_2015}. One alternative approach focuses on combining RCTs to estimate conditional average treatment effects in a non-parametric framework \citep{tan2022tree}. However, that work by Tan and colleagues was done in the federated learning setting, in which individual-level data could not be shared across study sites and instead only aggregate results or models could be shared. In the sections to follow, we tailor Tan et al.'s method to when individual-level data can be shared across trials, and we add other new options for combining trials.

To our knowledge, this paper is the first to describe and compare machine learning options for estimating heterogeneous treatment effects using data from multiple RCTs, in the setting in which all data can be shared across trials. Because not many methods exist to do this, we demonstrate several options for extending current methods for single studies to the multiple-study setting. We also build off of Tan et al.'s (\citeyear{tan2022tree}) approach by adapting it to the case when individual-level data can be shared across trials. Our goals are to assess estimation accuracy of the various methods within a given sample of trials and to determine whether and when pooling data is useful, or if it might ever worsen accuracy in the presence of high heterogeneity across trials. We conduct extensive simulations with varying data generation parameters to determine which of the single-study and aggregation methods perform best depending on different amounts of cross-trial heterogeneity in the effects. We then apply the approaches to a set of four RCTs of depression treatments and discuss the variability in estimates across the approaches and potential substantive conclusions that can be made.

\section{Notation}

The estimand considered in this paper is the conditional average treatment effect (CATE), defined under Rubin's potential outcomes framework \citep{rubin_estimating_1974}. Let $A$ denote a binary treatment indicator (often treatment versus control), $\boldsymbol{X}$ represent covariates, and $Y$ represent a continuous outcome. Under Rubin's framework, $Y(0)$ and $Y(1)$ denote the potential outcomes under control and treatment, respectively. In other words, $Y(0)$ is the value of $Y$ that an individual would have if they are in the control group, while $Y(1)$ is the value of $Y$ that they would have if they received treatment. The fundamental problem of causal inference is that we cannot ever observe both $Y(0)$ and $Y(1)$ simultaneously for the same person; therefore, we must use design and analysis approaches to estimate the unobserved outcomes. Next, let $S$ be a categorical variable representing the trial in which the individual participated and ranging from $1$ to $K$, where $K$ is the total number of RCTs. Finally, represent the probability of receiving treatment given covariates and trial membership (propensity score) as $\pi_s(\boldsymbol{X}) = P(A=1|\boldsymbol{X},S=s)$.

With a continuous outcome, the CATE is
\begin{equation} \label{unicate}
    \tau(\boldsymbol{X}) = E(Y(1)|\boldsymbol{X}) - E(Y(0)|\boldsymbol{X}).
\end{equation} In this paper, we note that the goal estimand is this ``universal'' CATE (\ref{unicate}) built off of potential outcomes that are not dependent upon study membership. However, many methods in the following sections target a study-specific CATE:

\begin{equation} \label{studycate}
    \tau_s(\boldsymbol{X}) = E(Y(1)|\boldsymbol{X},S=s) - E(Y(0)|\boldsymbol{X},S=s).
\end{equation} 

To identify the estimand when combining data across RCTs, many of the standard causal inference assumptions are required, including the Stable Unit Treatment Value Assumption (SUTVA) within each RCT. Other standard assumptions include: unconfoundedness (Assumption \ref{asx1}), consistency (Assumption \ref{asx2}) and positivity (Assumptions \ref{asx3} and \ref{asx4}) \citep{tan2022tree}. Assumption \ref{asx2} varies slightly depending on the estimand; under the universal CATE estimand (Equation \ref{unicate}), we assume overall consistency, while under the study-specific estimand (Equation \ref{studycate}), we assume consistency within each study. Assumption \ref{asx4}, which requires that any $\boldsymbol{X}$ is possible to be observed in all studies, can be relaxed depending on the method.

\newcommand{\indep}{\perp \!\!\! \perp}
\newtheorem{assumption}{Assumption}

\begin{assumption} \label{asx1} 
$\{Y(0), Y(1)\} \indep A \mid \boldsymbol{X}, S=s$~~for all studies $s$.
\end{assumption}
\begin{assumption} \label{asx2} 
$Y = AY(1) + (1-A)Y(0)$~~almost surely (in each study).
\end{assumption}
\begin{assumption} \label{asx3} There exists a constant $c>0$ such that $c<\pi_s(\boldsymbol{x})<1-c$ for all studies $s$ and for all $\boldsymbol{x}$ values in each study.
\end{assumption}
\begin{assumption} \label{asx4} (\textit{Can be relaxed})
There exists a constant $d>0$ such that $d<P(S = s|\boldsymbol{X}=\boldsymbol{x})<1-d$ for all $\boldsymbol{x}$ and $s$.
\end{assumption}



\section{Methods}

This paper includes methods developed for treatment effect estimation in a single study and aggregation approaches that apply these methods to multiple studies. This section discusses three single-study methods and several aggregation options that apply the single-study methods to the multi-study setting.


\subsection{Single-Study Methods} 
\label{SSL}

For a given RCT, many machine learning methods have been developed for CATE estimation. The single-study methods that exist can be grouped into multiple categories, as delineated by Brantner et al (\citeyear{brantner2023review}). For ease of comparison, three approaches are included that are user-friendly and have been shown to be effective in previous literature: the S-learner, X-learner \citep{kunzel2019metalearners}, and causal forest \citep{athey_generalized_2019}.  We ultimately selected these three approaches because they represent two distinct classes of methods for estimating heterogeneous treatment effects \citep{brantner2023review} and seem to be used in practice, especially the causal forest \citep{athey_estimating_2019, jawadekar2023practical}.  Specifically, the first two approaches are multi-step procedures that involve first estimating the conditional outcome mean under treatment or control and then combining the two into one CATE function, while the causal forest involves tree-based partitioning of the covariate space by treatment effect. In this paper, we use random forests as the base learners for both the S-learner and the X-learner to best compare with the causal forest, which is inherently forest-based. These single-study methods are different from those explored by Tan and colleagues (\citeyear{tan2022tree}); we chose to focus on the causal forest over a causal tree because the causal forest is an aggregation of multiple trees, and we added in the X-learner and S-learner to provide a different type of method to compare with.

\subsubsection{S-Learner}

The first single-study machine learning method used in this paper is called the "S-learner" \citep{kunzel2019metalearners}. This method is classified as a "meta-learner" in that it combines base learners (i.e., regression models) of any form in a specific way \citep{kunzel2019metalearners}. The S-learner uses a base learner (i.e., a random forest) to estimate a conditional outcome mean function given observed covariates and assigned treatment: $$\mu(\boldsymbol{X}, A) = E(Y|\boldsymbol{X}, A).$$ The conditional outcome mean function in this approach is not specific to treatment group, but instead treatment is included together with the covariates as features to be used by the random forest. The CATE can then be directly estimated by plugging in $0$ and $1$ for the treatment indicator to obtain predicted outcomes under treatment and control for each individual and calculate  $$\hat{\tau}(\boldsymbol{X}) = \hat{\mu}(\boldsymbol{X},1)-\hat{\mu}(\boldsymbol{X},0).$$

\subsubsection{X-Learner}

The second approach considered here is another meta-learner called the "X-learner" \citep{kunzel2019metalearners}. The X-learner takes a similar approach as the S-learner by modeling the conditional outcome mean functions before estimating the CATE directly. However, rather than estimating one outcome mean function for $Y(1)$ and $Y(0)$ simultaneously, the X-learner estimates two functions separately and then imputes treatment effects for each treatment group.

Specifically, the X-learner involves three steps. First, the conditional outcome mean functions are estimated using base learners (in this case, random forests) like in the S-learner, but separately by treatment group: $$\mu_0(\boldsymbol{X}) = E(Y(0)|\boldsymbol{X})~~~\text{and}~~~\mu_1(\boldsymbol{X}) = E(Y(1)|\boldsymbol{X}).$$ Next, the unobserved potential outcomes for individuals in the treatment and control groups are predicted using those models to get $\hat{\mu}_0(\boldsymbol{X}_{i: A=1})$ (estimate of the potential outcome under control for an individual who received treatment) and $\hat{\mu}_1(\boldsymbol{X}_{i: A=0})$ (estimate of the potential outcome under treatment for an individual who received control). We then input these predictions along with the observed outcomes to impute individual treatment effects: $$\Tilde{D}_{i: A=1} = Y_{i: A=1}-\hat{\mu}_0(\boldsymbol{X}_{i: A=1})~~~\text{and}~~~\Tilde{D}_{i: A=0} = \hat{\mu}_1(\boldsymbol{X}_{i: A=0})-Y_{i: A=0}.$$
Then $\Tilde{D}$ is regressed on $\boldsymbol{X}$ to estimate $\tau(\boldsymbol{X})$. This is done within each treatment group separately, resulting in two estimates, labeled $\hat{\tau}_1(\boldsymbol{X})$ and $\hat{\tau}_0(\boldsymbol{X})$. Finally, these are combined to obtain one estimate of the CATE function: $$\hat{\tau}(\boldsymbol{X}) = g(\boldsymbol{X})\hat{\tau}_1(\boldsymbol{X}) + (1-g(\boldsymbol{X}))\hat{\tau}_0(\boldsymbol{X}),$$ where the weight $g(\boldsymbol{X})$ is often an estimate of the propensity score (the case in this paper) or can be chosen otherwise \citep{kunzel2019metalearners}.

\subsubsection{Causal Forest} \label{cf}

The third single-study approach is the causal forest \citep{athey_generalized_2019}. The causal forest is similar to a random forest, but the focal estimand is the treatment effect itself, rather than the outcome for a given individual. The causal forest is based off of a causal tree, which involves recursive partitioning of the covariates to best split based on treatment effect heterogeneity. Here, the treatment effect is estimated as the difference in average outcomes between the treatment and control group individuals within leaves. From there, the causal forest is the weighted aggregation of many causal trees.

One potential challenge with causal forests is that bias could occur when there is overlap between the data used to form the trees and data used to estimate the treatment effects within leaves. A solution to that problem, called "honesty", has been proposed \citep{wager_estimation_2018}. This concept ensures that for every individual involved in creating the tree, their outcome is used either for splitting the tree or estimating the treatment effect within a leaf, but not both. Honesty has been used some in the literature, but there is not a widespread conclusion as to whether trees should be fit with or without honesty depending on the scenario. Dandl and colleagues compared honesty versus adaptive (not honest) forests in their simulations including causal forests and found that in their setting that was meant to represent an RCT, the adaptive forests performed better \citep{dandl2022makes}. Additionally, honesty requires large sample sizes. Thus, we do not include honesty in the causal forests in the primary simulations but do investigate it in a second round of method comparisons. 

\subsection{Aggregation Methods}

In many contexts, there are multiple RCTs available that compare the same two treatments. It is then worth considering methods that allow combining across trials. When aggregating to the multi-study level, the question becomes: how much does the treatment effect vary based on study membership? This variability can range along a continuum, where on one end is the possibility that the trials are all very homogeneous in terms of the CATE, meaning that participants in trial $j$ and in trial $k$ who have the same covariate values would have the same treatment effect. At the other extreme, individuals with the same covariates but in different trials could have completely different treatment effects. These differences can be due to heterogeneity in the sites in which the trials were conducted, heterogeneity in trial procedures (including the treatment or control conditions themselves), heterogeneity in trial samples, or other reasons. The aggregation methods to follow take different approaches to incorporating trial membership into the treatment effect estimation, ranging from assuming trial membership does not matter at all, to allowing it to matter just as much as any other characteristic.

\subsubsection{Complete Pooling}

A complete pooling approach is very straightforward: the researcher simply takes all data from each of the $K$ RCTs, creates a single dataset, and then fits one of the three previously described methods (S-learner, X-learner, or causal forest) to the pooled dataset. This approach is quick and easy to do, but requires many assumptions. Namely, this approach assumes a high level of homogeneity across trials and specifically that the CATE function is shared across studies. This method is included because it represents a naive comparison point and because it provides universal CATE estimates (i.e., not study-specific).

\subsubsection{Pooling with Trial Indicator}

An alternative pooling approach is to incorporate trial membership in the models but essentially still perform the pooling as before. Here, all of the individual data from each RCT is combined into one comprehensive dataset, but a categorical variable is included that represents the trial in which the individual participated. Then, the researcher can apply one of the single-study approaches to this full dataset, allowing for all of the covariates, including trial membership, to be involved in the treatment effect function. In this way, if trial membership is important for estimating effects, estimates should be somewhat informed by trial membership; otherwise, the treatment effect estimates should be similar across trials. While the previous complete pooling approach gives estimates that were not trial-specific, this approach yields trial-specific CATE estimates.

\subsubsection{Ensemble Approach}
The next approach is based off of Tan and colleagues' (\citeyear{tan2022tree}) methods for federated learning, originally developed for scenarios in which individual data cannot be shared across trial sites. Their original approach fits trial-specific models and then applies those models to data from a single coordinating site to derive an ensemble. We propose an adaptation of Tan's approach for settings where individual-level data from all trials are available to the analyst.

This adaptation of Tan et al.'s approach involves three steps. 
\begin{enumerate}
    \item First, the researcher builds localized models for the CATE within each trial, using one of the three single-study methods previously discussed (S-learner, X-learner, or causal forest).
    \item Next, they apply these localized models to each individual across all of the RCTs to get for each individual their \textit{trial-specific CATE estimates}, i.e., the estimated effects had the individual been part of study 1, study 2 and so on. For $K$ studies with a total of $N$ individuals in all studies combined, there will be $K$ trial-specific CATE models. Once each of these models are applied to all $N$ data points, every individual will have $K$ different estimates of their CATE. So there will ultimately be $N\cdot K$ CATE estimates in what Tan et al. define as an "augmented" dataset. The difference between the second step here and what Tan et al. did is that we apply the study-specific models to all data points in all trials, rather than having to restrict to a single coordinating site.
    \item The third and final step is to fit an ensemble model to the augmented dataset that has CATE estimates for every individual crossed with every trial. In this ensemble model, the response variable is the CATE estimate, and the predictors are the individual covariates and a categorical variable indicating the local model that had been used to compute the CATE estimate. We use three different options for this final ensemble model fit to the augmented dataset: a regression tree, a random forest, and a lasso regression. The regression tree and random forest were explored in Tan et al.'s (\citeyear{tan2022tree}) paper, while we added lasso regression to provide a parametric comparison point.
\end{enumerate} 

The resulting functions from these ensemble approaches are trial-specific estimates of the CATE; however, they have been adapted based on the CATEs from the other trials. Therefore, this method allows for trial heterogeneity but incorporates information across trials to hopefully improve the model from each trial. 

\subsubsection{IPD Meta-Analysis}

As a comparison point in the simulations to follow, we also include an individual patient-level data (IPD) meta-analysis with a random intercept for trial membership. This method is a standard approach taken by researchers when combining multiple RCTs and assessing treatment effects \citep{debray_get_2015, seo_comparing_2021, burke_meta-analysis_2017}, and it also serves here as a parametric comparison to the primarily non-parametric approaches outlined above. A meta-analysis does not employ a single-study method like the S-learner, X-learner, or causal forest; instead, all of the data is pooled together and trial-level relationships can be included as fixed or random effects. The decision of how to parametrize a given meta-analysis is very important and can have major implications as to the assumptions of how the true data is distributed and the subsequent fit of the model. While the previous non-parametric approaches implicitly allow for any important moderating relationships and interactions to be picked up based on the modeling procedure, meta-analysis requires that we pre-specify moderation according to a priori hypotheses. In this paper, we set up the meta-analysis to mimic the setup of the first scenario in the simulation to follow except for the exact form of the moderator, so that we can see how well meta-analysis performs when it is mostly correctly specified versus when it is incorrectly specified (for the second and third CATE scenarios described in the simulations below). The model is as follows:
$$Y = (\alpha_0 + a_s) + 
(\alpha_1+b_s)X_1 + \alpha_2X_2 + \alpha_3X_3 + \alpha_4X_4 + (\zeta + z_s) A + (\theta + t_s) X_1 A + \epsilon.$$ In this model, we allow the intercept to include a fixed component ($\alpha_0$) and a random component by study ($a_s \sim N(0, \sigma_a^2))$, and our residual error is $\epsilon \sim N(0, \sigma^2)$. The fixed effects are  $\boldsymbol{\alpha} = \{\alpha_1, \alpha_2, \alpha_3, \alpha_4\}$, the coefficients relating the covariates to the outcome; $\zeta$, the coefficient for treatment; and $\theta$, the coefficient of the interaction between treatment and a moderator $X_1$ \citep{seo_comparing_2021}. The random effects by study are $b_s \sim N(0, \sigma_b^2)$, the random slope for the covariate $X_1$; $z_s \sim N(0, \sigma_z^2)$, the random slope for treatment; and $t_s \sim N(0, \sigma_t^2)$, the random slope for the treatment-$X_1$ interaction term. From here, the estimate of the conditional average treatment effect can be calculated as $\hat{\tau}_s(\boldsymbol{X}) = (\hat{\zeta}+\hat{z}_s) + (\hat{\theta}+\hat{t}_s)X_1$.

The meta-analysis framework assumes that the CATE function is shared across studies, but that the mean potential outcome under control can differ across studies. Notably, this functional form of the CATE assumes linear relationships, and one must prespecify all variables that might be relevant to the main effect of the covariates and to the treatment effect.

\subsubsection{No Pooling}

Finally, we also consider that there might be instances where trials are too heterogeneous to reliably combine information across trials. When this is the case, fitting models within each study would be the best approach; therefore, we include this option in our simulations as well. For this ``no pooling'' approach, one can fit a single model within every trial separately using a single-study method previously introduced, and CATEs can be estimated within each study using the given study's model. We provide results from this method in the simulations to investigate if there are settings when pooling worsens estimation accuracy. However, it is important to mention that this approach is not technically an ``aggregation approach'' because it analyzes each study independently from the others and does not use data from multiple studies together. Particularly in the simulations to follow, the no pooling approach will find the best fit within each study and should therefore yield consistently high estimation accuracy. There will also be some differences in terms of variance; we assume that there would be higher variance when using only one study, but we do not explore this explicitly here. Note, though, that the current setup does not examine how well this approach will predict CATEs for individuals outside of the specific trials; we elaborate on this more in the sections to follow.


\section{Simulation Setup}

To compare both the single-study and aggregation methods, we performed a simulation study, simulating data from multiple randomized controlled trials and changing parameter values to compare which methods achieve the lowest mean squared error (MSE) between the estimated and true individual CATEs. Because there were three single-study methods (S-learner, X-learner, and causal forest) and six aggregation methods (complete pooling, pooling with trial indicator, ensemble tree, ensemble forest, ensemble lasso, and no pooling) being compared along with meta-analysis, there were $3\cdot 6+1=19$ total combinations of methods applied to each simulated dataset.

\subsection{Data Generating Mechanism}

In the simulations to follow, the potential outcomes are generated using the following model \citep{tan2022tree}: \begin{equation} \label{outcome}
    Y_i(a) = m(\boldsymbol{x}_i,s_i) + \frac{2a-1}{2}\cdot \tau(\boldsymbol{x}_i,s_i) + \epsilon_i
\end{equation} where $m(\boldsymbol{x}_i,s_i)$ represents the outcome mean conditional on covariates and trial, and $\tau(\boldsymbol{x}_i,s_i)$ is the CATE. In the main setting for the data generation, we employed two options for $m$ and $\tau$. The first setup (1a) involves a linear $m$ and piecewise linear $\tau$, based on a similar setup by Tan et al. (\citeyear{tan2022tree}): $$m(\boldsymbol{x},s) = x_1/2 + \sum_{j=2}^4x_j + \beta_s + \delta_s\cdot x_1 \text{   and   } \tau(\boldsymbol{x},s) = x_1\cdot I(x_1>0) + \beta_s + \delta_s\cdot x_1.$$ The second setup (1b) involves a more complicated non-linear function for $\tau$, derived partially from a simulation setting by Kunzel et al.  (\citeyear{kunzel2019metalearners}): $$m(\boldsymbol{x},s) = 0 \text{   and   } \tau(\boldsymbol{x},s) = g(x_1)g(x_2) + \beta_s + \delta_s\cdot x_1$$ where $g(x) = \frac{2}{1+\exp(-12(x-1/2))}$ \citep{kunzel2019metalearners}. In both of these, the coefficients $\beta_s$ represent trial-specific main effect coefficients, and $\delta_s$ represent trial-specific interaction effect coefficients (interaction between trial and the moderator $x_1$). In both setups, $x_1$ is an effect moderator, and in the second setup, $x_2$ is as well. If the coefficients $\beta_s$ and $\delta_s$ differ across $s$ (i.e., trial membership), then trial is making an impact in the moderation.

From this information, the components simulated are listed as follows:

\begin{enumerate}
    \item For each simulation, the number of trials was $K=10$.
    \item Each trial had a sample size of 500 individuals.
    \item Within each trial, we simulated five continuous covariates per person $X_i$, $i \in \{1,2,3,4,5\}$, where $E(X_i) = 0$, $Var(X_i) = 1$, and $Cov(X_i, X_j) = 0.2$ for all $i \neq j$.
    \item Each person was then assigned a treatment status, $0$ or $1$, according to a propensity score of $\pi_i = 0.5$ within each trial.
    \item Each person was also assigned an error term for their outcome function, so $\epsilon_i \sim N(0, 0.01)$.
    \item We then sampled trial-specific main effect and interaction effect terms. Each of the $K=10$ studies was assigned a main effect term according to $\beta_s \sim N(0, \sigma_{\beta}^2)$ and an interaction effect term according to $\delta_s \sim N(0, \sigma_{\delta}^2)$. The values of the standard deviations were: $(\sigma_{\beta}, \sigma_{\delta}) \in \{(0.5,0), (1,0), (1, 0.5), (1, 1), (3, 1)\}$.
    \item From this information, $m$, $\tau$, and $Y$ were calculated under either of the two setups described above (1a and 1b).
\end{enumerate}

We then included some variations of the above setup to assess method performance under different adjustments. The first was including one other scenario (2) to see how the methods would perform when the functional form of the CATE itself differed across trials -- a particularly challenging situation for pooling. For this scenario, we used the same form for $Y_i$ as in Equation (\ref{outcome}), and now we set $m$ and $\tau$ to be such that $m$ is linear and $\tau$ depends on study: $$m(\boldsymbol{x},s) = x_1/2 + \sum_{j=2}^4x_j$$ and $$\tau(\boldsymbol{x},s) = I(s \in \{1,2,3,4\})\cdot g(x_1)g(x_2) + I(s \in \{5,6,7,8\})\cdot x_1\cdot I(x_1>0) + I(s \in \{9,10\})\cdot 0$$ where $g(x)$ is as previously defined.

We also added settings with variation in the trial sample sizes. One new option involved one large trial (n=1000) and the rest smaller (n=200).  The second new setting had half of the trials with n=500 and the other half with n=200. We assessed performance for these sample size adjustments under scenarios 1a, 1b, and 2 with trial main and interaction coefficient standard deviations of 1 and 0.5, respectively. 

We then investigated the impact of covariate shift on method performance. In particular, we generated the data such that all even numbered studies had $X_1$ with mean 0 as above, but in odd numbered studies, the mean of $X_1$ was set to be 2. We assessed this setting under scenarios 1a and 1b with standard deviations of 1 and 0.5 of study main and interaction effect terms, and we allowed trial sample sizes to either all be the same or for one trial to be large and the rest smaller.

Finally, we added some simulations with $K=30$ trials in two of the settings (scenario 1a and 1b with trial main and interaction coefficient standard deviations of 0.5 and 0, respectively) to determine if there were differences in performance based on number of trials (the remainder of the simulations had $K=10$). 

For each simulation setup, we generated 1,000 simulated datasets. Necessary packages included causalToolbox for the S-learner and X-learner \citep{kunzel2019metalearners}, grf for the causal forest \citep{athey_generalized_2019}, rpart for the ensemble tree \citep{therneau2015package}, ranger for the ensemble forest \citep{wright_ranger_2017}, glmnet for the ensemble lasso \citep{friedman2017package}, and lme4 for the mixed effects meta-analysis \citep{bates2010lme4}. Ensembling functions were based off of those in the ifedtree package \citep{tan2022tree} but were adapted to the setting in which data could be shared across trials. In all non-parametric approaches, hyperparameters were set to be the defaults, except that the causal forest was set to use 1,000 trees instead of the default of 2,000 for computational ease, and honesty was set to false for the preliminary simulations. For each method and each iteration, performance of the different approaches was assessed based on the mean squared error (MSE) between the true individual CATEs and the estimated individual CATEs, and these MSEs were ultimately averaged across the 1,000 repetitions. Code containing all adapted methods and implementation of the simulations can be found at the github repo: \verb|https://github.com/carlyls/CATE_multiRCT|.

\section{Simulation Results}

\begin{figure}[h!]
    \centering
    \includegraphics[width=18cm]{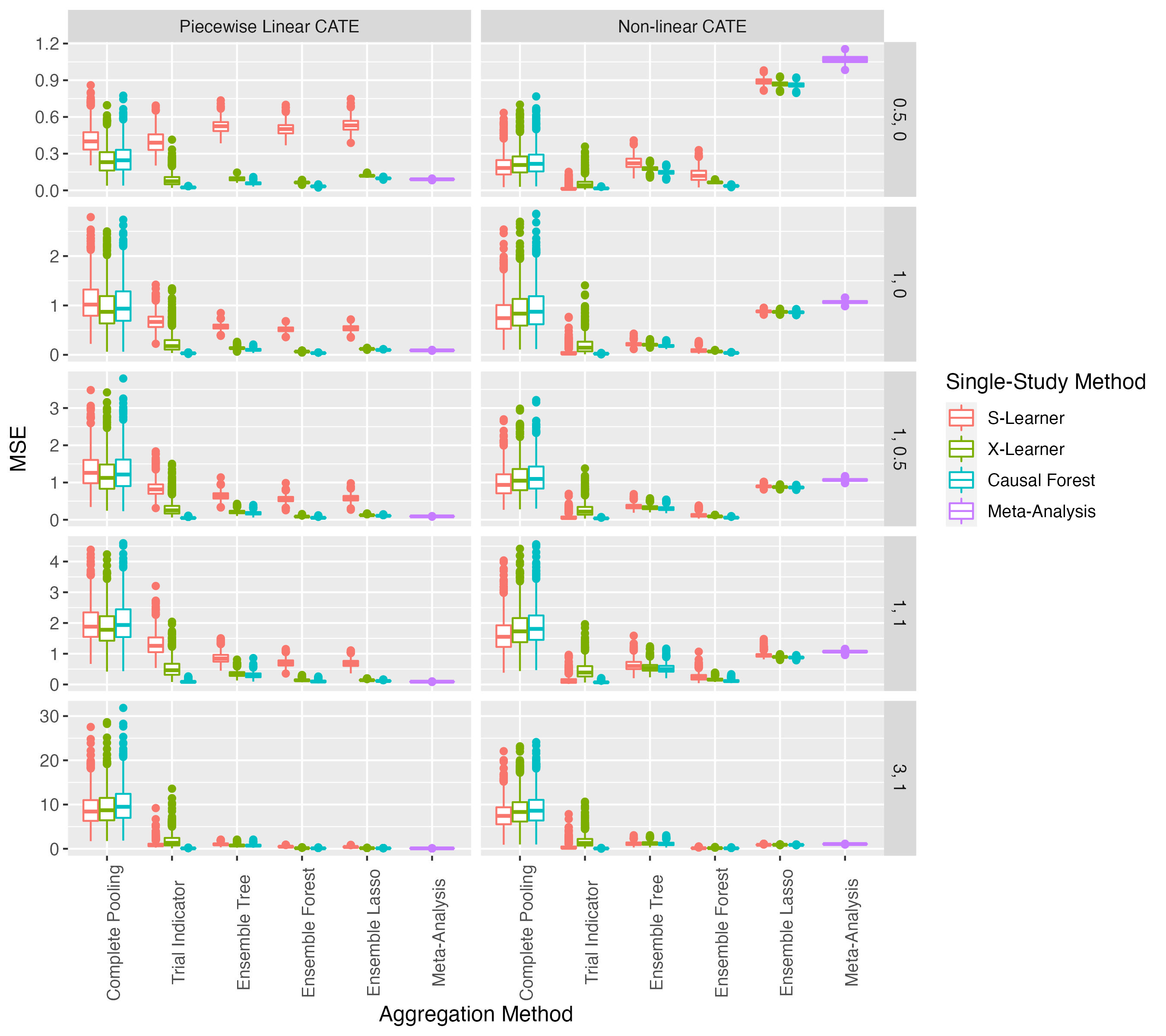}
    \caption{Distribution of MSE for main parameter combinations across all single-study and aggregation approaches.*\\
    \footnotesize *Columns are broken down by simulation scenarios (piecewise linear versus non-linear CATE), and rows are by standard deviation of study main and study interaction coefficients.}
    \label{fig1}
\end{figure}

The following tables and figures display the performance results across 1,000 iterations of each parameter combination/scenario. Figure \ref{fig1} displays the distribution of MSE for every approach for the two main scenarios (piecewise linear and non-linear CATE), broken down by the standard deviations of the trial main and interaction effects. In the piecewise linear and non-linear CATE scenarios, as the trial coefficients (both main and interaction effects) increase in variability, the MSE increases, meaning the methods estimate individual CATEs more poorly. This is consistent with the idea that when trial membership is involved in the treatment effect function, the CATEs vary across trials and therefore are harder to estimate when data is pooled across studies. Notably, this increase in MSE happens much more quickly for the complete pooling approaches.

In the piecewise linear scenario (1a), the most consistently effective approaches in terms of MSE are when the causal forest is used as the single-study method and when the aggregation approach is either pooling with trial indicator or ensemble forest. The X-learner also performs relatively well in terms of MSE. Meta-analysis performs well, which is expected because the model was set up to mostly match the true functional form of the CATE in this scenario. For the non-linear scenario (1b), the ensemble lasso and meta-analysis perform notably worse, which makes sense due to the complexity of the functional form of the CATE, as it includes the product of two expit functions, and the lasso and meta-analysis assume a parametric linear relationship between covariates and outcome. The ensemble forest and pooling with trial indicator again estimate the CATEs well, with all single-study methods performing more similarly. While the S-learner was not very effective with the piecewise linear CATE (1a), it was more effective with the non-linear CATE (1b). In all main settings, the no pooling approach performs similarly well to pooling with trial indicator and ensemble forest (Figure A\ref{nopool}); we discuss more about this in the Discussion section.

Several boxplots in the Appendix display the results of the many variations upon the original simulation settings included. To assess the performance of methods with different trial sizes, Figure A\ref{samplesize} demonstrates that there do not seem to be notable differences in patterns across methods depending on whether all trials have the same sample size, one trial is much larger, or half are larger while half are smaller. The MSE seems to be slightly higher overall when trial sizes are different, but not substantially different. Furthermore, Figure A\ref{sc2} displays the results for the variable CATE scenario (2). Here, the causal forest is clearly performing the best of the three single-study methods, while the S-learner is not performing as well. The most effective aggregation methods are again pooling with trial indicator and ensemble forests, and meta-analysis performs relatively poorly.

When we introduced a shift in the covariate distributions between even versus odd numbered studies (Figure A\ref{covshift}), there again does not seem to be a difference in the patterns of results. The MSE generally is slightly higher across all methods compared to when the covariates all came from the same distributions across trials; however, methods like the causal forest with pooling with trial indicator and ensemble forest still perform consistently well. In the piecewise linear CATE with a shift in covariate distributions, meta-analysis performs very well and the best of all aggregation approaches, but it does not perform well when the CATE is non-linear. 

Finally, for the two scenarios with 30 trials instead of 10, Figure A\ref{k30} demonstrates that the results and patterns are all similar to the results for K=10, except for the causal forest with pooling with trial indicator. Interestingly, this approach, which performed very well with 10 trials, has high MSE when there are 30. To understand this more fully we did further investigations, including some iterations with 15, 20, and 25 trials to see how the pattern changes. Overall, the results of these investigations indicate that when there are more trials, the causal forest with pooling with trial indicators has more difficulty identifying the heterogeneity that exists across trials. In particular, the method rarely ``picks up'' the trial indicators of trials that do have different patterns in effects when $K > 20$, as indicated by the variable importance measures (weighted sum of the number of times the variable was used in a split at each level of the forest) \citep{athey_generalized_2019}. Table A\ref{ktest} shows average variable importance values under the piecewise linear CATE scenario for different values of $K$. Based on the simulation setup, the causal forests should split often on moderating variables, which in this case are $X_1$ and study membership. The variable importance measures demonstrate that for all values of $K$, $X_1$ is involved in a high proportion of splits, as it should be as a moderator. For lower numbers of trials ($K = 10$ through around 20), the most heterogeneous studies (defined based on main coefficients) had notable variable importance, meaning they were involved in some of the splits in the causal forest. However, for higher values of $K$ (more trials), the variable importance for these most heterogeneous studies approached zero, meaning study membership was no longer picked up much in the causal forest even though there was notable heterogeneity of the treatment effect based on study membership. In addition, for high values of $K$, the causal forest split more often on the non-moderating covariates, $X_2 - X_5$. These issues that arose with large numbers of trials likely contributed to the high MSE of the causal forest with pooling with trial indicator for large values of $K$. We reflect more on these results in the Discussion section.

To more formally examine the results of the main settings in our simulation, we regressed the average MSE across iterations on the methods and parameter combinations, just within the piecewise linear and non-linear CATE scenarios, excluding meta-analysis and no pooling, and excluding the settings with $K=30$ and with covariate shift. Specifically, the regression is such that: \begin{align*}
   MSE =  &\beta_0 + \beta_1\cdot singlestudy + \beta_2\cdot aggregation + \beta_3\cdot singlestudy\cdot aggregation \\
   &+ \beta_4\cdot main_{sd} + \beta_5\cdot interaction_{sd} + \beta_6\cdot scenario + \beta_7\cdot trialsizes + \epsilon.
\end{align*} From this regression, there were no significant differences in performance across single-study methods, but all aggregation methods performed significantly better than complete pooling. The ensemble forest had the best average MSE for the S-learner and X-learner, and pooling with trial indicator had the best average MSE for the causal forest. 


Finally, we also performed 500 more iterations using the same methods previously described, but with honest causal forests instead of traditional ``adaptive'' causal forests. These iterations were performed using the main data generation setups as above, except that covariates were not correlated. The resulting average MSEs are presented in the Appendix (Figure A\ref{ahonest}). We found very similar results to the original 1,000 repetitions with adaptive causal forests, but the honest causal forests had slightly higher MSEs on average, indicating worse estimation accuracy than the adaptive causal forests. For the ensemble tree, forest, and lasso, the honest causal forests had slightly higher average MSE compared to the X-learner (Figure A\ref{ahonest}), while the adaptive causal forests had slightly lower average MSE compared to the X-learner for these same aggregation approaches in the original simulations. However, these differences are very small, so we can broadly make similar conclusions whether we use adaptive or honest causal forests in these scenarios.

\section{Application to Real Dataset}

After the simulations demonstrated differences across methods in several data generation setups, we applied the various methods to an existing dataset containing multiple randomized controlled trials that compared the same two medications.

\subsection{Treatments for Major Depressive Disorder}

The applied dataset used in the current paper consists of four randomized controlled trials \citep{mahableshwarkar_randomized_2013,mahableshwarkar_randomized_2015,boulenger_efficacy_2014,baldwin2012randomised}, each of which included three treatments: duloxetine, vortioxetine, and placebo, where duloxetine and vortioxetine are both treatments for major depressive disorder (MDD). At the time of the trials, duloxetine had been more commonly used to treat MDD so was primarily included in the trials as an active reference, while vortioxetine was a newer treatment not yet marketed \citep{schatzberg2014overview}. Each of the four trials compared at least two different dosages of vortioxetine and therefore had more participants taking vortioxetine as opposed to duloxetine or placebo. For the purposes of the current application, we removed placebo participants and lumped all dosages of vortioxetine together to investigate the potential differences between the efficacy of the active medications (duloxetine and vortioxetine), as well as identify features that might be moderating this difference.

Participants in each of the four trials shared similar eligibility criteria. All four trials required patients to be between the ages of 18 to 75, to have a Major Depressive Episode (MDE) as a primary diagnosis according to the DSM-IV-TR criteria over at least three months, and to have a Montgomery-Asberg Depression Rating Scale (MADRS) \citep{montgomery1979new} score of at least 22 (one trial) or 26 (three trials) at both screening and baseline \citep{mahableshwarkar_randomized_2013,mahableshwarkar_randomized_2015,boulenger_efficacy_2014,baldwin2012randomised}. A primary outcome in the trials is the change in MADRS (Montgomery-Asberg Depression Rating Scale) \citep{montgomery1979new} score from baseline to the last observed follow-up in the study. Participants were meant to stay in the study for 8 weeks, at which point their final MADRS score was collected. For those who did not remain in the trial for 8 weeks, a last observation carried forward imputation approach was used for simplicity. This imputation approach is not the best way to account for missing data and many other options exist \citep{little2012prevention}, but it is used here for simplicity because this example is primarily illustrative. Predictors/effect modifiers used in the models were age, sex (female or male), smoking status (ever smoked or never smoked), weight, baseline MADRS score, baseline HAM-A (Hamilton Anxiety Rating) score \citep{hamilton1959assessment}, comorbidity indicators (if ever had diabetes mellitus, hypothyroidism, anxiety), and medication indicators (if they are concomittantly taking an antidepressant, antipsychotic, thyroid medication). Since the outcome is the difference in MADRS score (MADRS at follow up minus MADRS at baseline), a more negative outcome indicates a better result. We removed individuals who were either in the placebo group or who had missing treatment assignment, along with individuals with missing baseline MADRS or no post-randomization MADRS. After this, sample sizes were 575, 436, 418, and 418 for each of the trials. Further descriptive information about the samples in the four RCTs is reported in the Appendix (Table A\ref{desc}). Little missing covariate data was present in the sample; however, conditional mean imputation was performed for missing values of weight (n=1) and baseline HAM-A score (n=2).

Following data preparation, we used each of the aforementioned method combinations (i.e., causal forest, S-learner, and X-learner as single-study methods paired with complete pooling, pooling with trial indicator, ensemble tree, ensemble forest, and ensemble lasso) to estimate the CATEs for every individual across the four trials. We then compared the CATE estimates across methods to see their concordance levels. Notably, it is not possible to compare the method performances with the truth, as the true CATEs are unknown in this real dataset.

\subsection{Results}

All methods broadly led to the conclusion of a positive average CATE. This indicates that vortioxetine is estimated to have less of a beneficial effect on the MADRS score on average. In each of the four RCTs, both treatments were associated with a reduction in depressive symptom severity over time (shown through a reduction in MADRS score), but this reduction was smaller for the vortioxetine group than the duloxetine group. Table \ref{cates} contains the mean and standard deviation of the CATEs according to each method. Broadly, the S-learner approaches estimated lower CATEs on average than the other approaches, and there is some consistency between the aggregation approaches within each single-study method (S-learner, X-learner, and causal forest). There were especially high levels of similarity in the average CATE estimates across the causal forest methods, shown in the last column of Table \ref{cates}. The variability of the CATE estimates differs depending on the approach as well; causal forest approaches had higher standard deviations than approaches that used the S-learner and X-learner. Complete pooling also yielded the highest standard deviations for CATE estimates out of all of the aggregation approaches. As a comparison point, we used a multiple linear regression with a random effect for trial to estimate an average treatment effect of 2.49 (SE = 0.49), which is similar to the averages of the CATEs according to the X-learner and causal forest approaches.

\begin{figure}[h!]
    \centering
    \includegraphics[width=12cm]{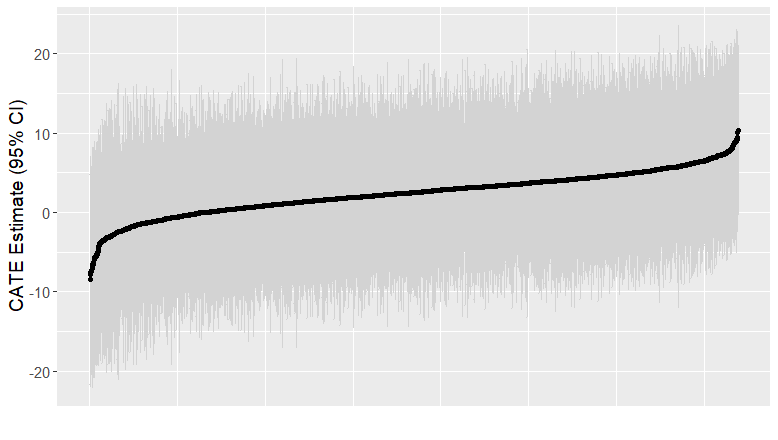}
    \caption{Point estimates and 95\% confidence intervals for CATEs according to causal forest with pooling with trial indicator.}
    \label{catecis}
\end{figure}

We then focused in on results from the causal forest with pooling with trial indicator approach, since that approach performed the best on average in the simulations when there were not a large number of trials being combined. The CATE estimates and their 95\% confidence intervals from this approach are displayed in Figure \ref{catecis}. These confidence intervals were calculated based on variance estimates provided through the grf package, where variance is calculated based on comparison of individual CATE predictions within and across small groups of fitted causal trees \citep{athey_generalized_2019}. These estimates support that the majority of individuals have a positive CATE estimate, but they also display very high levels of uncertainty, with all confidence intervals including zero. 

\begin{figure}[h!]
    \centering
    \begin{subfigure}[t]{0.49\textwidth}
        \centering
        \includegraphics[width=\linewidth]{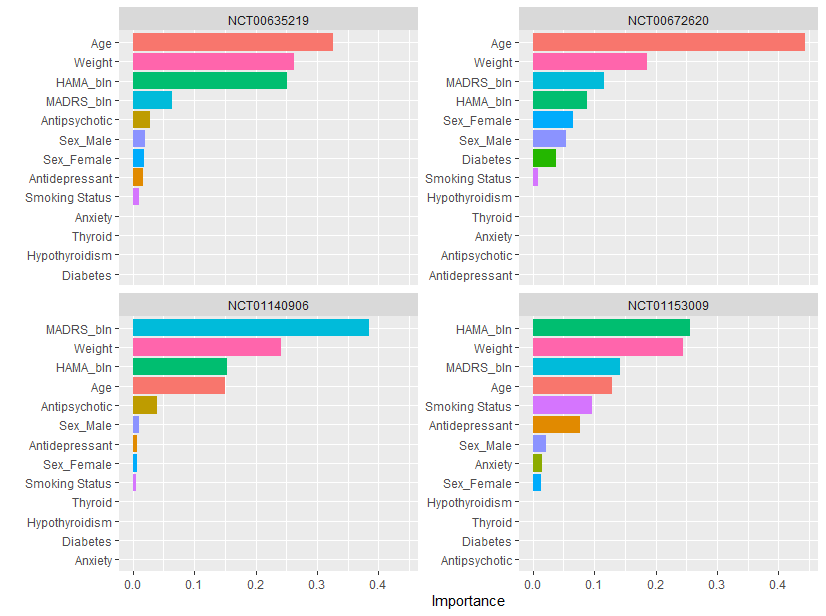} 
        \caption{Variable importance for study-specific causal forest models.} \label{studyspec_varimp}
    \end{subfigure}
    \hfill
    \begin{subfigure}[t]{0.49\textwidth}
        \centering
        \includegraphics[width=\linewidth]{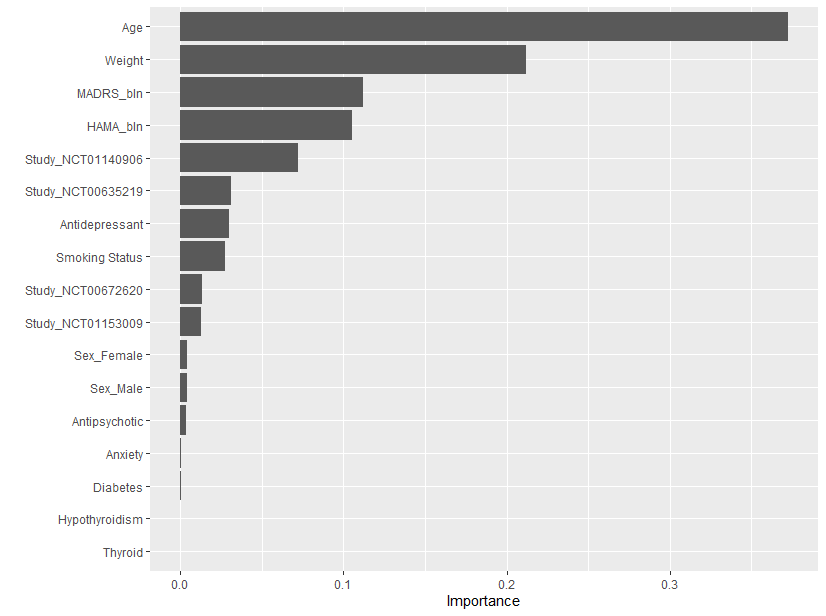} 
        \caption{Variable importance for causal forest with pooling with trial indicator} \label{cfpool_varimp}
    \end{subfigure}
    \caption{Variable importance measures (a) within studies, and (b) according to the causal forest with pooling with trial indicator}
    \label{varimp}
\end{figure}

To learn more about the moderation within the CATE model, we can explore variable importance measures. As previously mentioned, variable importance from the grf package \citep{athey_generalized_2019} is a weighted sum of the number of times the variable was used in a split at each level of the forest. Figure \ref{varimp} displays the variable importance measures according to the grf package \citep{athey_generalized_2019} for all covariates, first in separate causal forests for each study (\ref{studyspec_varimp}), and second according to the causal forest with pooling with trial indicator (\ref{cfpool_varimp}). From Figure \ref{studyspec_varimp}, there are a few variables that are consistently identified as effect moderators across studies (age, weight, baseline MADRS score, and baseline HAM-A score), and there are several that are not found to be major moderators (the comorbidity and medication indicators). However, notably there are some differences according to the separate models, indicating that the treatment effect functions are slightly different within each study. Figure \ref{cfpool_varimp} then displays the resulting importance measures from one aggregation model fit to all studies. Here, we can see that the same four variables (age, weight, baseline MADRS, and baseline HAM-A) are involved in a high proportion of the splits in the causal forest, and study membership is involved in some splits as well. The fact that these study indicators are not more highly involved in the partitioning of the treatment effect is a good sign, though, that there is not a very high level of heterogeneity in CATE estimates across studies. 

\begin{figure}[h!]
    \centering
    \includegraphics[width=13cm]{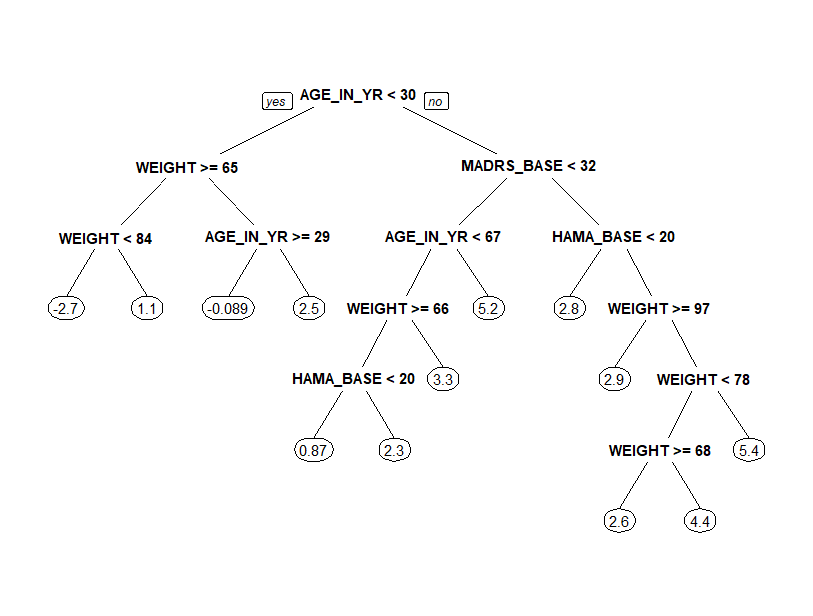}
    \caption{Interpretation tree for causal forest with pooling with trial indicator.*\\
    \footnotesize *Circled numbers represent the average CATE estimate for individuals in that leaf.}
     \label{inttree}
\end{figure}

The variable importance plots do not demonstrate the direction of the moderating effect, however. We briefly investigate these directional effects through an interpretation tree (Figure \ref{inttree}) and through exploratory plots such as Figure A\ref{aage}. This interpretation tree was formed by fitting a regression tree, where the CATE estimates according to the causal forest with pooling with trial indicator were the outcomes, and the features (predictors) were every covariate in the original CATE model. The tree confirms what was shown in Figure \ref{varimp} -- that age, weight, baseline MADRS, and baseline HAM-A score are the strongest effect moderators. Study membership does not show up in this interpretation tree, supporting that there is low heterogeneity across trials. This is a helpful visual to see the direction of the relationships aggregated across trials, but it is exploratory and should not be interpreted in great detail. Another similar approach for investigating the CATE function in terms of individual moderators is to fit the best linear projection of the CATE estimates using a function in the grf package \citep{athey_generalized_2019}; the resulting coefficients from this regression using doubly-robust estimates of the CATE are reported in Table A\ref{atab2}.

Broadly, these interpretations of the CATE function derived by the causal forest with pooling with trial indicator do not display high levels of heterogeneity, with the exception of potential heterogeneity by age. The scatterplot of CATE estimates by age in Figure A\ref{aage} and the best linear projection summarized in Table A\ref{atab2} indicate somewhat higher CATE estimates for older individuals; however, there are very high levels of uncertainty in the confidence intervals (Figure \ref{catecis}). Other than this potential moderation by age, there does not appear to be heterogeneity across other variables, and in general we suggest further study, perhaps using more trials or observational data, to assess whether this age relationship is truly strong.

We also can compare the results of these pooled non-parametric methods with a more standard approach -- IPD meta-analysis. In particular, we fit a linear regression with random effects for trial membership and included interaction terms to investigate potential moderation and compare results to the causal forest with pooling with trial indicator. As previously mentioned, the IPD meta-analysis yielded an average treatment effect estimate of 2.49 (SE = 0.49). To go a step further, we added interaction terms between treatment and each covariate in separate models to determine whether any interaction terms were significant. None were, although the interaction for age was close to significant (95\% CI: (-0.01, 0.14)), which is consistent with our findings in the non-parametric approaches. We finally performed a subgroup analysis where we divided the sample into four groups based on age (18-34, 35-44, 45-54, and 55-75 years old) and fit mixed effects regression models with random effects for trial membership to each subgroup separately. The resulting average treatment effect estimates for each subgroup are presented in Figure A\ref{subgroup}, and they lead to a similar conclusion -- that older individuals may have a higher treatment effect, but the moderation does not appear to be statistically significant.

This data application shows how to effectively apply the methods compared in simulations to a real dataset and assess potential moderation. The methods all agree broadly on the direction of the average treatment effect but imply somewhat different conclusions with respect to the individual CATE estimates. In comparing the causal forest with pooling with trial indicator versus the IPD meta-analysis with trial random effects, we reach similar conclusions. We expand upon the benefits and drawbacks of these approaches in the following section.


\section{Discussion}

In this paper, we compared methods to estimate the conditional average treatment effect in a single trial and methods to extend the single-trial approaches to multiple trials. In the absence of notable cross-trial heterogeneity of treatment effects, the methods examined all performed well, but when trial membership was involved in the treatment effect function, some methods performed worse than others. Specifically, and not surprisingly, methods that ignore trial membership (complete pooling) do not effectively estimate the CATE when there is cross-trial heterogeneity. On the other hand, some methods performed well no matter the level of heterogeneity: pooling with trial indicator and ensemble forests had consistently low mean squared error despite increasing the variability of the trial membership coefficients in the treatment effect. This was especially true when the single-study method used was the causal forest (Figure \ref{fig1}). These patterns held across various data generation setups, including introducing different sample sizes across trials and a covariate shift. The patterns persisted for the most part with 30 trials as opposed to 10; however, the causal forest with pooling with trial indicator performed much worse with 30 trials. Therefore, this approach could be highly effective with a smaller number of trials but might miss key study-level differences with a large number of trials. Having 30 trials to combine is unlikely in practice, though, in our experience. Otherwise, the two best performing methods -- causal forest with pooling with trial indicator and causal forest with ensemble forest -- showed high accuracy across all other scenarios and could be good first choices for combining trials to estimate heterogeneous treatment effects.

When considering the three single-study approaches, the most consistently favorable method in the simulations was the causal forest, followed by the X-learner. The S-learner performed well in certain scenarios, such as scenario 1b, where the treatment effect function involved a bounded, non-linear expit function. The performances of the S-learner and X-learner in our simulations and applied example were consistent with results found previously \citep{kunzel2019metalearners}, in that the S-learner seemed to be somewhat biased towards 0 in the applied example (Table \ref{cates}) and performed worse in the simulations when the treatment effect function was complicated (variable CATE scenario and the piecewise linear and non-linear CATE with high variability). The X-learner performed well in the simulations with complex CATEs and with structural forms of the CATE, again consistent with previous work \citep{kunzel2019metalearners}. The causal forest performed well across all scenarios. These simulation results and the results from the applied data example of MDD medications demonstrate that it is important to carefully select the single-study method for a given question, as each of the three options can provide different estimates. A good starting point would be to consider expert knowledge of how heterogeneous across studies and complicated the outcomes or treatment effect might be. These results also indicate the need for more diagnostics to help researchers determine which approach to choose. In general though, the causal forest performed consistently well when combining 10 studies, so use of this method is supported by the simulations.

The simulations also incorporated some comparisons between the non-parametric and parametric approaches. Specifically, the use of a lasso regression as an ensemble showed how a parametric ensemble could perform compared to the ensemble tree and forest. The lasso performed very well when the treatment effect function was piecewise linear (scenario 1a) but quickly suffered in performance when the function was more non-linear (scenarios 1b and 2). Furthermore, the inclusion of a mixed effects meta-analysis demonstrated a common parametric technique used in the multiple-study setting. This model was set up to perform well when the CATE function was piecewise linear (scenario 1a), but it yielded high MSE in the non-linear and complex scenarios that it was not correctly parametrized for (scenario 1b and 2). The particular specification of a meta-analysis is therefore very important, and incorrect hypotheses of key interactions and moderating relationships have major implications for model fit and accuracy of estimates. These comparisons demonstrate that non-parametric machine learning approaches are very beneficial when the treatment effect function is complicated and non-linear, as the non-parametric methods do not require correct specification of any parameters. Although interpretability becomes more of a challenge, the non-parametric methods allow for flexible relationships and hopefully high levels of accuracy in CATE estimation. 

In this work, we did not explore an exhaustive list of potential single-study and aggregation methods, and we also investigated a few data generation setups that do not cover every possible scenario of real data. We attempted to select single-study methods that were common, user-friendly, and shown to be effective or potentially effective in previous literature. However, as this is an ever-growing field, future work could include other single-study methods \citep{wendling_comparing_2018,powers_methods_2018} to see how they compare to the ones used in this study. For example, it would be interesting to investigate the performance of the X-learner with a different base learner, such as Bayesian additive regression trees (BART) \citep{chipman2010bart,kunzel2019metalearners}. In general, non-parametric methods for CATE estimation are notably flexible and effective in estimating complex functional forms of the CATE; however, reliable variance estimation for these approaches is somewhat lacking. Without the distributional assumptions present in parametric methods, the non-parametric approaches often require resampling procedures to effectively estimate variance in predictions. Furthermore, with ensemble approaches such as those used in this paper, there are multiple sources of variance coming from both the original predictions and the predictions from the ensemble model. Therefore, variance estimation is an important area of future work for many of the methods discussed in this paper. 

Another important point related to the non-parametric approaches used in this work is that they primarily serve to accurately estimate the true CATE function. They are not as straightforward to use when the goal is identification of key moderators; although we can use tools like variable importance, there are not statistical tests of moderation as there are in parametric approaches like meta-analysis. In the simulations, we were thus not able to efficiently evaluate the methods' ability to identify effect moderators and instead prioritized minimizing error in CATE estimation. If a research goal is to identify moderators, some of the more exploratory work in the applied example (plotting CATE estimates, best linear projections, etc.) could be a helpful starting point, and potential moderators could then also be included in a parametric model to more formally test for moderation.

The approaches discussed in this paper implicitly rely on the assumption that all of the trials being combined have observed the same covariates $X$ necessary to estimate the CATE. We did not discuss cases where the trials contain different measures of a similar construct or cases of systematic missingness, meaning where certain covariates are not at all available in some trials. Approaches for dealing with systematic missingness have been discussed in the literature \citep{audigier2018multiple, jolani2015imputation} but not in this specific context, so future work should explore methods for addressing missingness and discordant measures of similar constructs. 

It is important to note that with the exception of complete pooling, the resulting CATE estimates are trial-specific. Unless trial was not picked up in the aggregation methods, the majority of the methods discussed will produce trial-specific estimates of the CATE. This allows for improved accuracy of estimates but might be less helpful in real world applications. We are interested in continuing to identify ways in which researchers could aggregate across trials to develop estimates that are accurate but not trial-specific -- this could be crucial for use of the resulting methods and models in practice, on data not coming from the specific trials used in the model formulation. However, the trial-specific estimates can still be useful; for example, if trials were done in separate hospitals, CATEs of future patients could be predicted using the hospital that they are being treated in, and the model that estimates their treatment effect should be more accurate after taking into consideration the data from the other hospitals. Similarly, the focal point of this paper and the simulations described above were the performance of models in the given sample. We thus assessed the performance in the simulations based on MSE across the trials used to fit the model, and we discuss accuracy in terms of the trials themselves. Future work will be focused on assessing how these methods perform when estimating CATEs in a target population, outside the specific trials used to estimate the CATE. This is where we might see even more of the benefits of pooling/ensembling approaches over methods like the no pooling approach, because we would be gaining information by combining trials.

In the MDD trials, duloxetine was included as a reference medication because it was already marketed at the time of the trials, and patients were excluded from the study if they had previously not responded to duloxetine. On the other hand, vortioxetine was not yet marketed and was the more experimental medication; therefore, some bias could arise due to participants being excluded if they had previously not responded to duloxetine. Acknowledging this, we were able to estimate treatment effects according to each method combination, and we used variable importance and interpretation trees to investigate which variables might be important moderators of the treatment effect. Variable importance is a limited measure and can often be biased towards continuous variables with more possible split points \citep{strobl2007bias}, so we encourage caution when interpreting those results. This example dataset shows how to combine multiple RCTs to get an improved assessment of treatment effect heterogeneity and better determine which treatment would be best suited to a given individual, based on their features and their estimated CATE.  Notably, the four trials used in this dataset were run by the same organizations and had very similar protocols; this helps ensure that we can confidently combine datasets but also might limit the potential heterogeneity across trials that might exist in other applications. We also did not see high levels of heterogeneity in the treatment effects based on other covariates in these trials. A general idea is that studies need to be four times larger to identify effect moderators compared to an average treatment effect \citep{fleiss2011design}, and this study included precisely four trials. Therefore, our findings would become more robust and we could more confidently assess heterogeneity or lack thereof with the inclusion of more studies.

There are many openings for future work, some of which have been mentioned. Broadly, it is important to further refine these methods and identify which are most helpful in specific data scenarios. It will also be helpful to determine when it is appropriate to develop universal CATE estimates, versus when the CATE estimates should be trial-specific. This paper demonstrated several approaches that take data from multiple studies and estimate heterogeneous treatment effects, using flexible models that allow for complex relationships -- which is often the case in the real world.

\section{Acknowledgments}

The study was funded by the Patient-Centered Outcomes Research Institute (PCORI) through PCORI Award ME-2020C3-21145 (PI: Stuart) and the National Institute of Mental Health (NIMH) through Award R01MH126856 (PI: Stuart). Ms. Brantner also received financial support in the form of a training grant through the National Institutes of Health (T32AG000247). Disclaimer: Opinions and information in this content are those of the study authors and do not necessarily represent the views of PCORI or NIMH. Accordingly, PCORI and NIMH cannot make any guarantees with respect to the accuracy or reliability of the information and data.

Furthermore, this paper is based on research using data from data contributors, Takeda and Lundbeck, that has been made available through Vivli, Inc. Vivli has not contributed to or approved, and is not in any way responsible for, the contents of this publication. This study, carried out under YODA Project 2022-4854, used data obtained from the Yale University Open Data Access Project, which has an agreement with Janssen Research \& Development, L.L.C. The interpretation and reporting of research using this data are solely the responsibility of the authors and does not necessarily represent the official views of the Yale University Open Data Access Project or Janssen Research \& Development, L.L.C.

\bibliographystyle{unsrtnat}
\bibliography{MLSims_Arxiv}

\newpage

\begin{table}[h!]
    \centering
    \begin{tabular}{|l|c|c|c|}
         \hline
         & \textbf{S-Learner} & \textbf{X-Learner} & \textbf{Causal Forest}\\
         \hline
         \textbf{Complete Pooling} & 1.38 (1.6) & 2.57 (1.4) & 2.37 (2.8) \\
         \textbf{Pooling with Trial Indicator} & 0.91 (1.3) & 2.52 (1.3) & 2.37 (2.7)\\
         \textbf{Ensemble Tree} & 0.89 (1.3) & 2.35 (1.5) & 2.23 (2.5)\\
         \textbf{Ensemble Forest} & 0.89 (1.1) & 2.36 (1.4) & 2.30 (2.2)\\
         \textbf{Ensemble Lasso} & 0.89 (1.2) & 2.32 (1.4) & 2.23 (2.1)\\
         \hline
    \end{tabular}
    \caption{Mean (SD) of CATEs from all individuals in sample according to different single-study and aggregation method combinations.*\\
    \footnotesize *The CATEs are individual-level estimates that indicate the difference in the estimated effect of vortioxetine versus duloxetine on the difference in MADRS score for a given patient. A positive CATE indicates that vortioxetine is estimated to have a smaller reduction of the MADRS score.}
    \label{cates}
\end{table}

\newpage
\newpage

\appendix

\begin{figure}[htp]
    \centering
    \includegraphics[width=14cm]{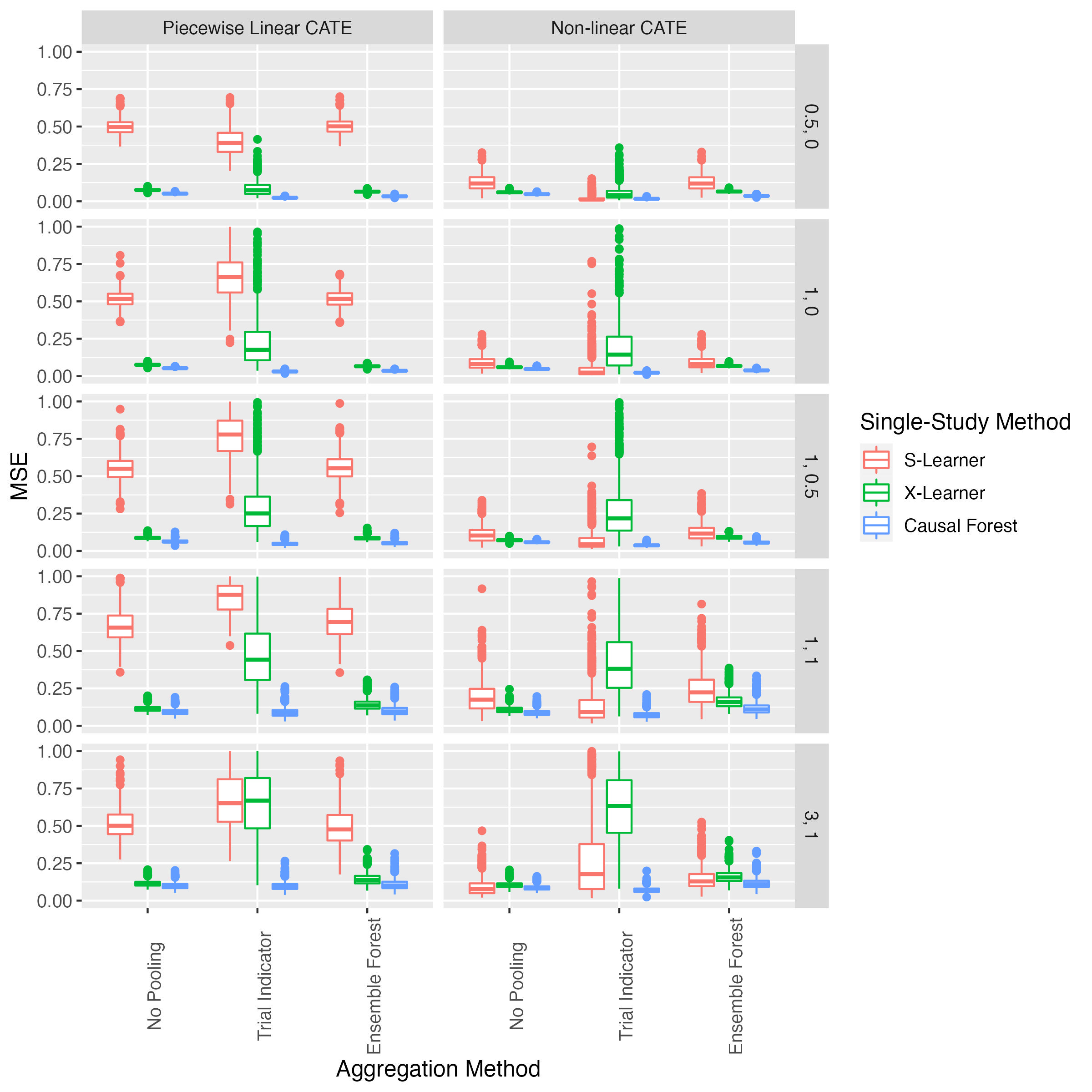}
    \caption{Distribution of MSE for no pooling versus best performing pooling/ensembling methods*\\
    \footnotesize *Columns are broken down by simulation scenarios (piecewise linear versus non-linear CATE), and rows are by standard deviation of study main and study interaction coefficients. Y-axis is cutoff for ease of visualization.}
    \label{nopool}
\end{figure}

\begin{figure}[htp]
    \centering
    \includegraphics[width=18cm]{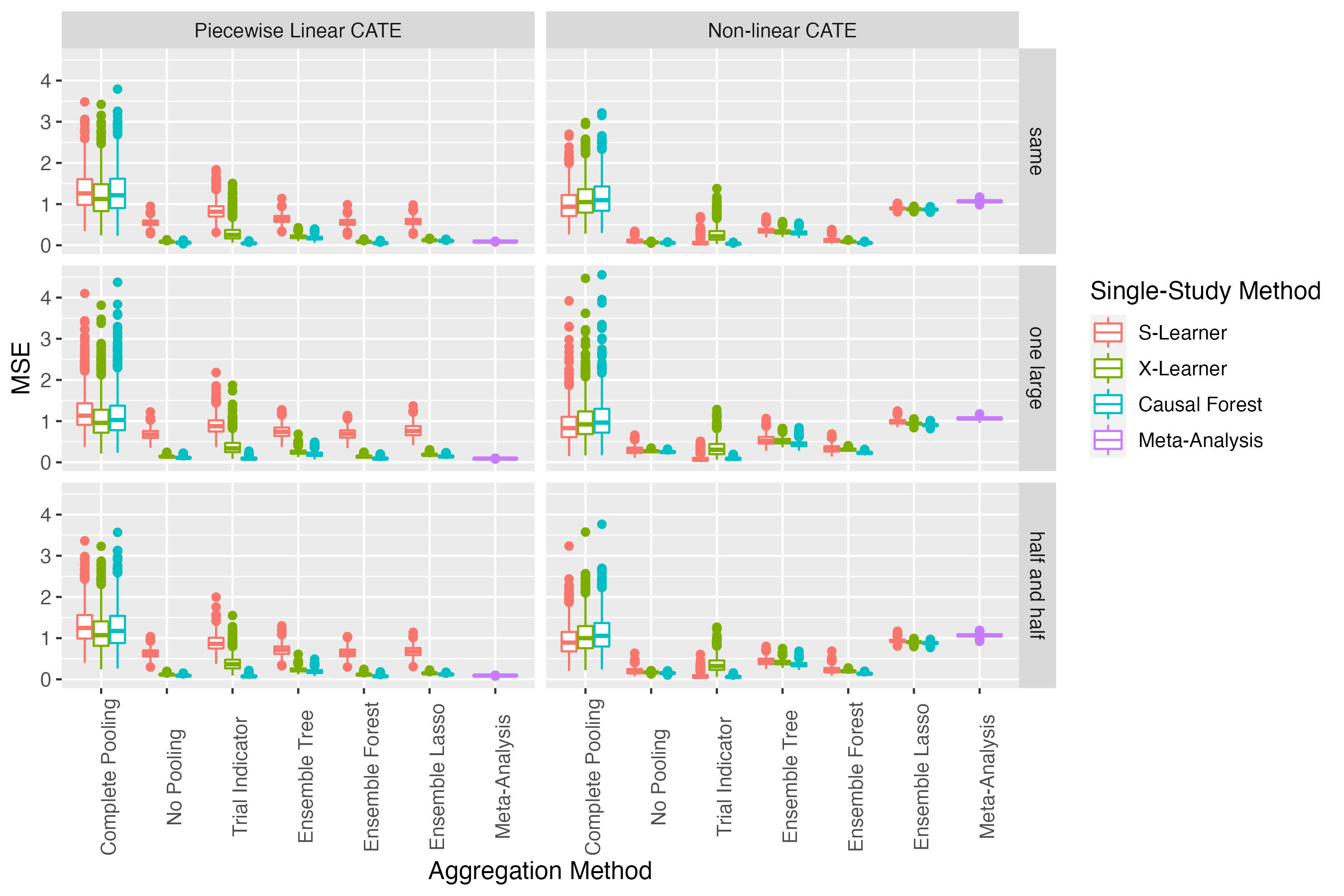}
    \caption{Distribution of MSE for trials with different sample sizes*\\
    \footnotesize *Columns are broken down by simulation scenarios (piecewise linear versus non-linear CATE), and rows are by trial sample sizes (same: all trials with n=500, one large: one trial with n=1,000 and the rest with n=200, half and half: five trials with n=500 and five with n=200). SD of study main and study interaction coefficients were 1 and 0.5, respectively for all iterations.}
    \label{samplesize}
\end{figure}

\begin{figure}[htp]
    \centering
    \includegraphics[width=18cm]{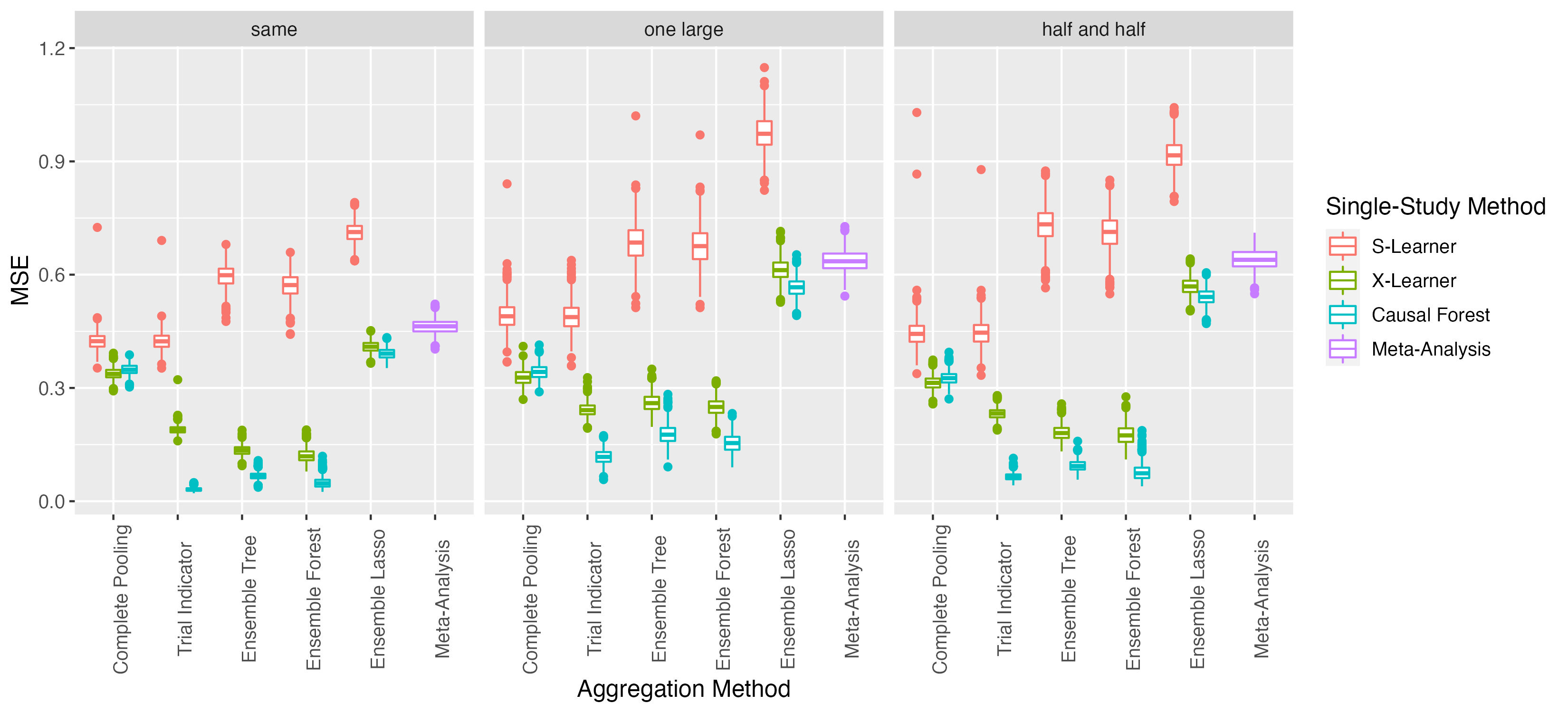}
    \caption{Distribution of MSE for trials with variable CATE function*\\
    \footnotesize *Columns are broken down by trial sample sizes (same: all trials with n=500, one large: one trial with n=1,000 and the rest with n=200,  half and half: five trials with n=500 and five with n=200). SD of study main and study interaction coefficients were 1 and 0.5, respectively for all iterations.}
    \label{sc2}
\end{figure}

\begin{figure}[htp]
    \centering
    \includegraphics[width=18cm]{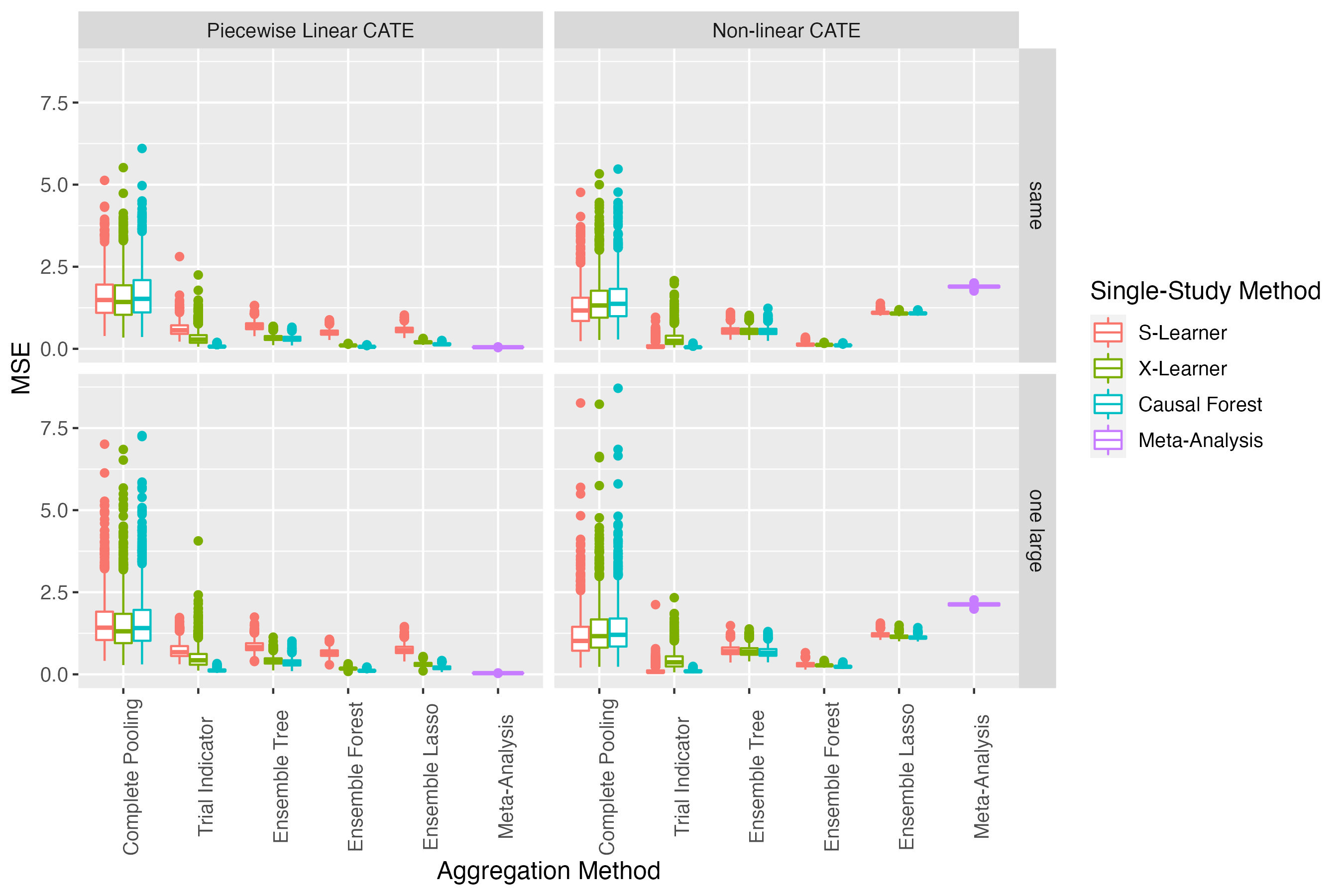}
    \caption{Distribution of MSE for trials with covariate shift*\\
    \footnotesize *Columns are broken down by simulation scenarios (piecewise linear versus non-linear CATE), and rows are by trial sample sizes (same: all trials with n=500, one large: one trial with n=1,000 and the rest with n=200). SD of study main and study interaction coefficients were 1 and 0.5, respectively for all iterations.}
    \label{covshift}
\end{figure}

\begin{figure}[htp]
    \centering
    \includegraphics[width=18cm]{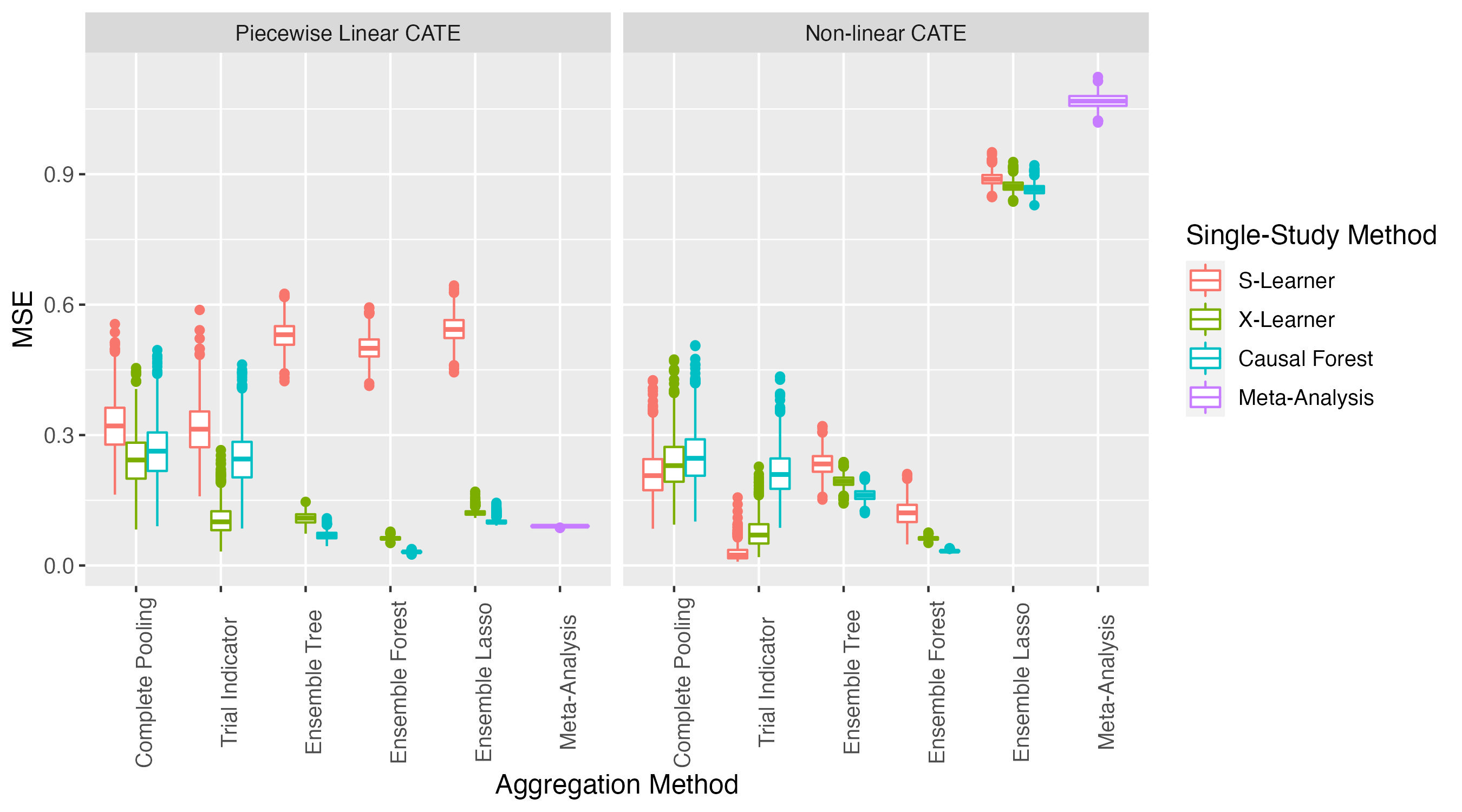}
    \caption{Distribution of MSE for K=30 trials*\\
    \footnotesize *Columns are broken down by simulation scenarios (piecewise linear versus non-linear CATE). SD of study main and study interaction coefficients were 0.5 and 0, respectively for all iterations.}
    \label{k30}
\end{figure}

\begin{table}[h!]
    \centering
    \begin{tabular}{|l|c|c|c|c|c|c|c|c}
    \hline
         \textbf{$K$} & \makecell{\textbf{Avg Importance}\\ \textbf{for} $X_1$ \\ Mean (SD)} & \makecell{\textbf{Avg Importance} \\ \textbf{for} $X_2-X_5$ \\ Mean (SD)} & \makecell{\textbf{Avg Importance for 20\% } \\ \textbf{Most Heterogeneous Studies}  \\ Mean (SD)} & \makecell{\textbf{Largest Absolute Value} \\ \textbf{Study Main Coefficient}  \\ Mean (SD)}\\
         \hline
         10 & 0.757 (0.07) & 0.008 (<0.01) & 0.090 (0.04) & 0.906 (0.24) \\
         \hline
         15 & 0.783 (0.02) & 0.006 (<0.01) & 0.059 (0.01) & 1.002 (0.25) \\
         \hline
         20 & 0.819 (0.02) & 0.007 (<0.01) & 0.034 (0.01) & 1.013 (0.21)\\
         \hline
         25 & 0.807 (0.01) & 0.044 (0.02) & 0.002 (<0.01) & 1.105 (0.22) \\
         \hline
         30 & 0.722 (0.01) & 0.069 (0.03) & <0.001 (<0.01) & 1.178 (0.23) \\
         \hline
    \end{tabular}
    \caption{Average variable importance measures across 50 iterations of causal forest with pooling with trial indicator for different values of K (the number of trials).*\\
    \footnotesize*Data is generated under a setting of a piecewise linear CATE with all trials the same size (n=500), no covariate shift, a study main coefficient standard deviation of 0.5, and a study interaction coefficient standard deviation of 0. Numbers reported represent average and standard deviations of variable importance measures according to the causal forest. The top 20\% of studies refer to the studies that had coefficients that were furthest in absolute value from the mean coefficient across all studies, meaning studies that had the most heterogeneity of the treatment effect.}
    \label{ktest}
\end{table}

\begin{figure}[htp]
    \centering
    \includegraphics[width=13cm]{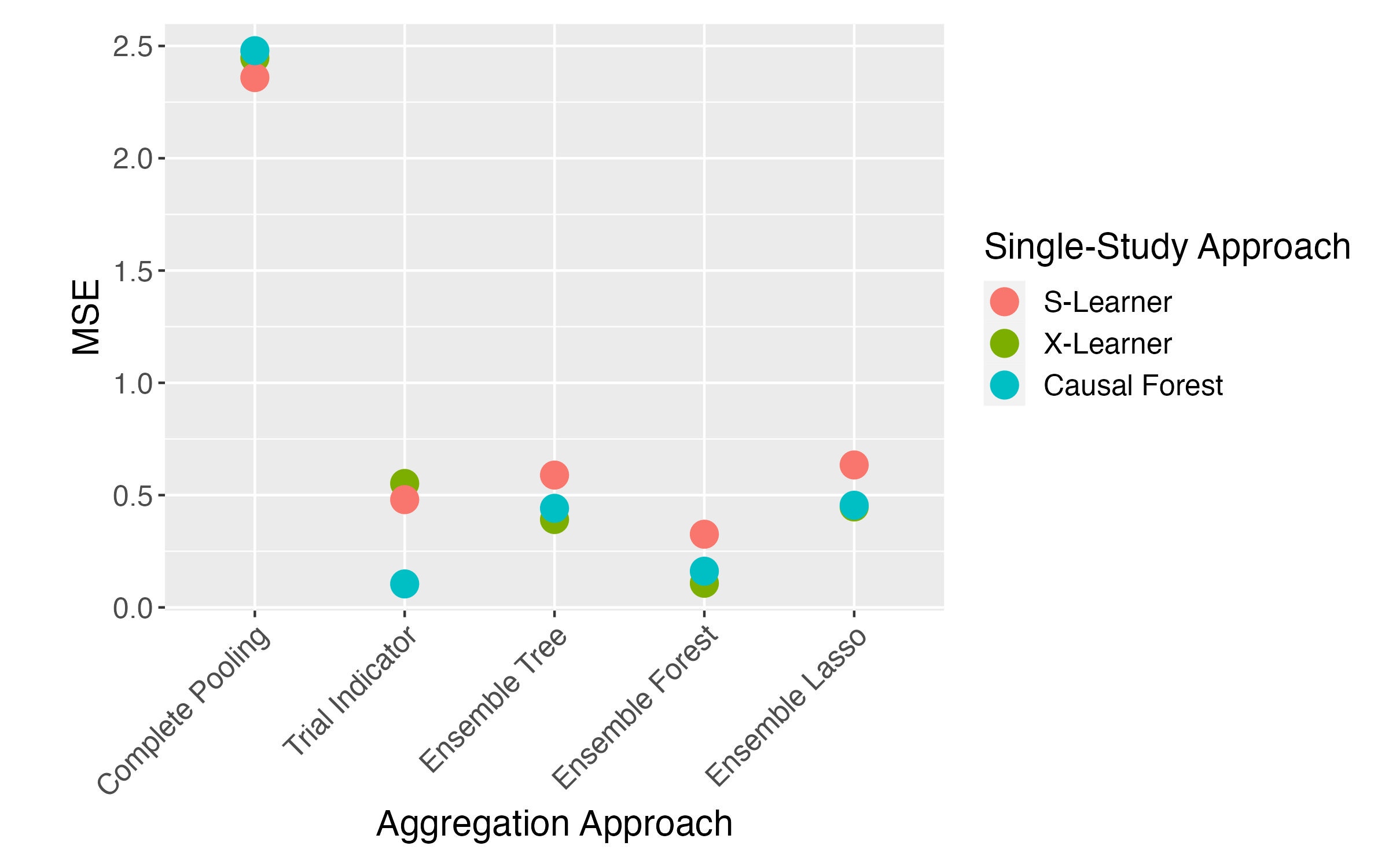}
    \caption{Average MSE across all scenarios and iterations using honest causal forests.}
    \label{ahonest}
\end{figure}

\begin{table}[h]
    \centering
    \includegraphics[width=18cm]{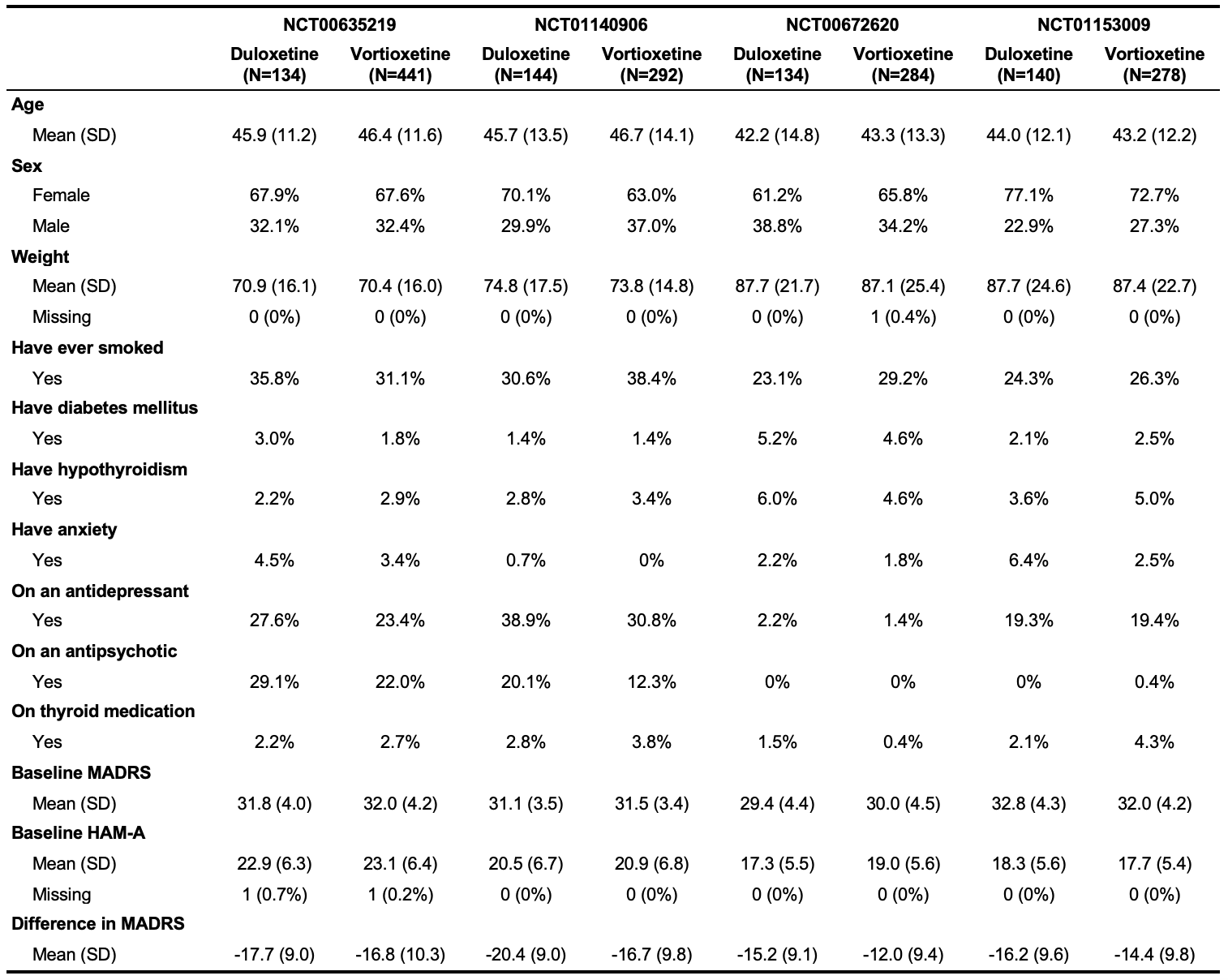}
    \caption{Descriptive statistics of participants of four randomized controlled trials, broken down by treatment group.}
    \label{desc}
\end{table}


\begin{figure}[h]
    \centering
    \includegraphics[width=13cm]{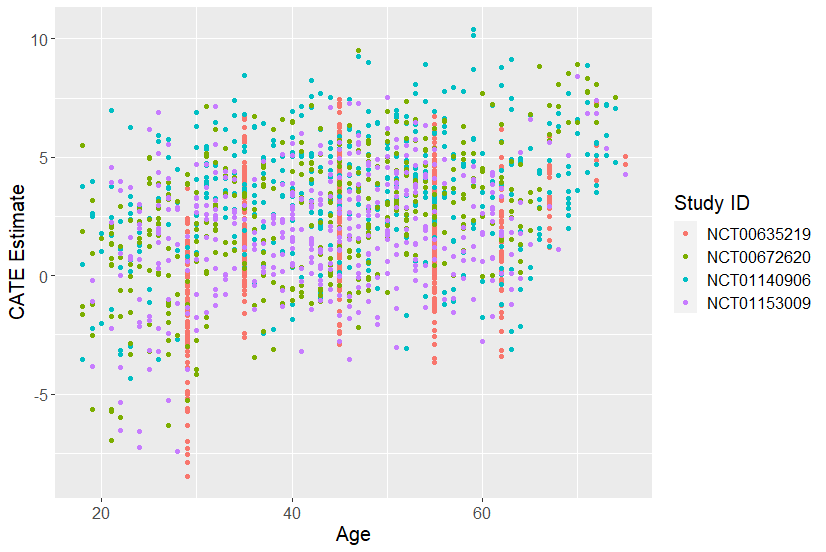}
    \caption{CATE estimates by age of individual according to causal forest with pooling with trial indicator.*\\
    \footnotesize *Note that uncertainty of the CATE estimates is not reflected in this plot.}
    \label{aage}
\end{figure}


\begin{table}[h]
    \centering
    \begin{tabular}{|l|c|c|c|}
    \hline
          & \textbf{Estimate} & \textbf{Standard Error} & \textbf{P-Value} \\
          \hline
         (Intercept) & -6.32 & 5.06 & 0.21 \\
         \hline
         Age & 0.09 & 0.04 & 0.03* \\
         \hline
         Female & 0.45 & 1.07 & 0.67 \\
         \hline
         Smoker & -1.47 & 1.10 & 0.18 \\
         \hline
         Weight & -0.01 & 0.03 & 0.72 \\
         \hline
         Baseline MADRS & 0.09 & 0.13 & 0.49 \\
         \hline
         Baseline HAM-A & 0.08 & 0.09 & 0.38 \\
         \hline
         Has Diabetes Mellitus & -3.97 & 3.36 & 0.24 \\
         \hline
         Has Hypothyroidism & -1.81 & 3.56 & 0.61 \\
         \hline
         Has Anxiety & 3.58 & 3.63 & 0.32 \\
         \hline
         Takes Antidepressant & 1.55 & 1.32 & 0.24 \\
         \hline
         Takes Antipsychotic & -0.21 & 1.93 & 0.91 \\
         \hline
         Takes Thyroid Medication & 2.23 & 4.22 & 0.60 \\
         \hline
         Study NCT00635219 & -1.09 & 1.63 & 0.50 \\
         \hline
         Study NCT01140906 & 2.71 & 1.62 & 0.09 \\
         \hline
         Study NCT00672620 & 2.93 & 1.58 & 0.06 \\
         \hline
    \end{tabular}
    \caption{Results of best linear projection of the CATE according to the causal forest with pooling with trial indicator.\\
    \footnotesize *Indicates a p-value less than 0.05.}
    \label{atab2}
\end{table}

\begin{figure}
    \centering
    \includegraphics[width=13cm]{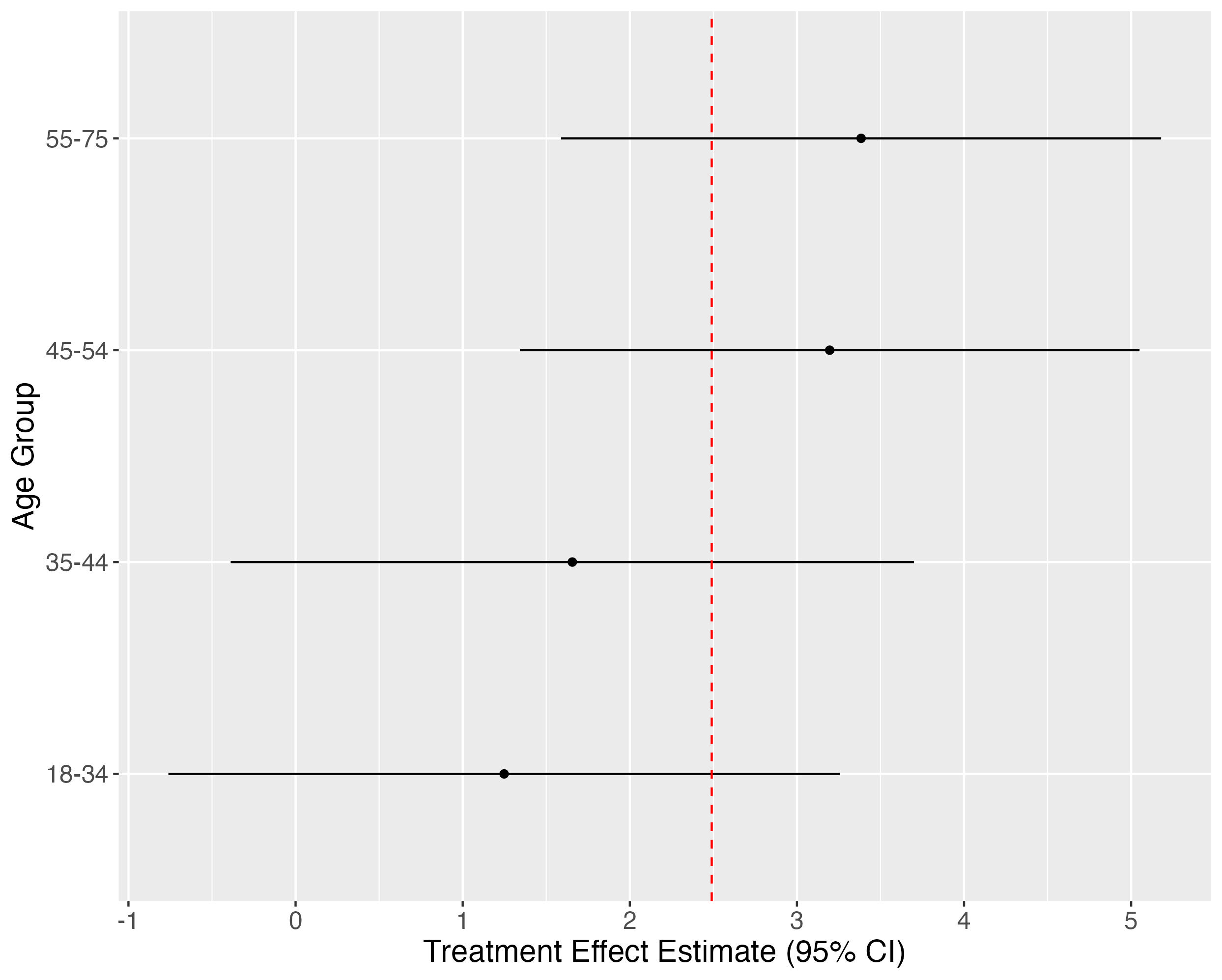}
    \caption{Average treatment effect with 95\% confidence interval by subgroup of age*\\
    \footnotesize*Vertical red line represents the overall average treatment effect estimate.}
    \label{subgroup}
\end{figure}

\end{document}